\documentclass[12pt,dvips]{article}

\usepackage{amsmath,amssymb,amsfonts}
\usepackage{cite}
\usepackage{axodraw}
\usepackage{graphicx}
\usepackage{bbm}

 \columnwidth\textwidth

\newcommand{\qR}{q_\text{\tiny$R$}}
\newcommand{\qbR}{\overline{q}_\text{\tiny$R$}}
\newcommand{\qL}{q_\text{\tiny$L$}}
\newcommand{\qbL}{\overline{q}_\text{\tiny$L$}}

\newcommand{\bm}{\mf}         
\newcommand{\nn}{\nonumber}
\newcommand{\slsh}[1]{\mbox{$\not\! #1$}}
\def\mf#1{\text{\boldmath{$#1$}}}

\newcommand{\lf}{\left}
\newcommand{\rg}{\right}

\date{\empty}	
\title{       
  {\normalsize
    \hfill DESY 04-204 \\[-1ex]
    \hfill HD-THEP-04-45} \\
  \vspace{1cm}
  \textbf{Factorisation, Parton Entanglement and the Drell-Yan Process}\\
  [8mm]}
      
\author{D.~Boer\,$^{a,1}$, A.~Brandenburg\,$^{b,2}$,
O.~Nachtmann\,$^{c,3}$ and 
  A. Utermann\,$^{c,4}$\\
  \\[2ex]
  {\normalsize
    $^a$ Dept.\ of Physics and Astronomy, Vrije Universiteit Amsterdam}
  \\
  {\normalsize De Boelelaan 1081, 1081 HV Amsterdam, 
The Netherlands}
\\[1ex]
   {\normalsize $^b$ Deutsches Elektronen-Synchrotron DESY}\\
 {\normalsize  Notkestra\ss e 85, D-22603 Hamburg, Germany}
\\[1ex]
  {\normalsize
    $^c$ Institut f\"ur Theoretische Physik, Universit\"at Heidelberg}\\
  {\normalsize Philosophenweg 16, D-69120 Heidelberg, Germany
}}
                                                          
\begin{document}
\begin{titlepage} 
  \maketitle
\vfill
\begin{abstract}
\noindent
We discuss the angular distribution of the lepton pair in the
Drell-Yan process, $\text{hadron}+\text{hadron}\to\gamma^\ast+X\to
l^++l^-+X$. This process gives information on the spin-density matrix 
$\rho^{(q,\bar{q})}$ of the annihilating quark-antiquark pair in
\mbox{$q+\bar{q}\to l^++l^-$}. There is strong experimental evidence that even
for unpolarised initial hadrons $\rho^{(q,\bar{q})}$ is nontrivial,
and therefore the quark-antiquark system is polarised. We discuss the
possibilities of a general $\rho^{(q,\bar{q})}$ --which could be
entangled-- and a factorising $\rho^{(q,\bar{q})}$. We argue that
instantons may lead to a nontrivial $\rho^{(q,\bar{q})}$ of the type
indicated by experiments.
\end{abstract}
\thispagestyle{empty}
\vspace{3ex}
\hrule width 5.cm
\vspace*{.5em}
{\small \noindent $^1$ email: dboer@nat.vu.nl \\ $^2$ email:
Arnd.Brandenburg@desy.de \\ $^3$ email:
O.Nachtmann@thphys.uni-heidelberg.de \\ $^4$ email:
A.Utermann@thphys.uni-heidelberg.de }
\end{titlepage}
\newpage \setcounter{page}{2}

\section{Introduction}\label{intro}
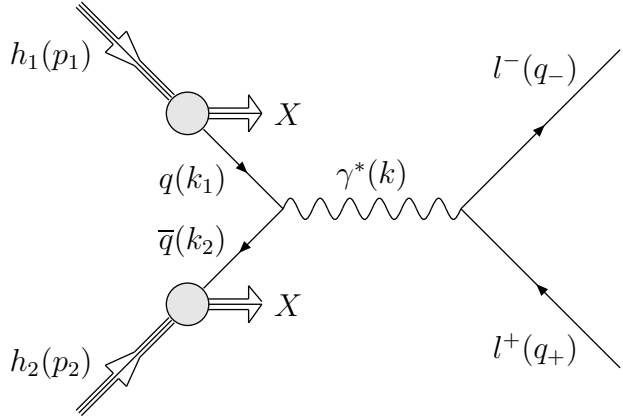
\begin{figure}
\centering
\begin{picture}(230,156)(0,0) 
 \Line(26,1)(61,36)
 
 \Line(27,0)(41,14)
 \Line(41,14)(44,11)
 \Line(44,11)(49,22)
 \Line(49,22)(62,35)

\Line(25,2)(39,16)
 \Line(39,16)(36,19)
 \Line(36,19)(47,24)
 \Line(47,24)(60,37)

\Text(0,20)[l]{$h_2(p_2)$}

 \Line(67,42)(95,42)
 \Line(67,40)(90,40)
 \Line(67,44)(90,44)

\Text(100,42)[lc]{$X$}

 \Line(90,44)(90,49)
 \Line(90,49)(95,43)
 \Line(95,42)(90,35)
 \Line(90,35)(90,40)

 \GCirc(67,42){8}{0.9}
 \ArrowLine(103,78)(73,48)
 \Text(82,66)[rc]{$\overline{q}(k_2)$}
 \ArrowLine(73,108)(103,78)
 \Text(82,89)[rc]{$q(k_1)$}
 \Photon(103,78)(170,78){4}{6} 
 \Text(137,85)[cb]{$\gamma^\ast(k)$}
 \ArrowLine(170,78)(230,138)
 \Text(215,131)[rc]{$l^-(q_-)$}
 \ArrowLine(230,18)(170,78)
 \Text(215,25)[rc]{$l^+(q_+)$}
 \Line(26,155)(61,120)
 \Line(27,156)(41,142)
 \Line(25,154)(39,140)

\Line(41,142)(44,145)
\Line(39,140)(36,137)

\Line(44,145)(49,134)
\Line(36,137)(47,132)

\Line(60,119)(47,132)
\Line(62,121)(49,134)

\Text(0,136)[l]{$h_1(p_1)$}

 \Line(67,114)(95,114)
 \Line(67,112)(90,112)
 \Line(67,116)(90,116)

\Text(100,114)[lc]{$X$}

 \Line(90,116)(90,121)
 \Line(90,121)(95,114)
 \Line(95,114)(90,107)
 \Line(90,107)(90,112)

 \GCirc(67,114){8}{0.9}

\end{picture}
\caption{\small\label{fig1} 
 The generic Drell-Yan process.}
\end{figure}

\noindent
In this note we discuss the question of factorisation in the Drell-Yan process \cite{dy}
\begin{eqnarray}\label{1}
&&h_1(p_1)+h_2(p_2)\rightarrow
\gamma^*(k)+X\nonumber\\
&&\hspace{3.7cm}\hookrightarrow l^+(q_+)+l^-(q_-)\,.
\end{eqnarray}
Here, $h_1$ and $h_2$ are the initial hadrons, $\gamma^*$ is the virtual photon,
$l^+,l^-$ are the final state leptons $(l=e,\mu)$ and $X$ stands for the hadronic final state
particles. The four-momenta are indicated in brackets. The basic underlying process
is the annihilation of a quark-antiquark pair
\begin{equation}\label{2}
q(k_1)+\bar{q}(k_2)\rightarrow\gamma^*(k)
\rightarrow l^+(q_+)+l^-(q_-)\,,
\end{equation}
which is sketched in Fig.~\ref{fig1}. Here we focus on the
discussion of reaction (\ref{2}) which is the lowest order process in
the framework of the QCD improved parton model, see for
instance~\cite{alta}. For massless quarks we find that in (\ref{2}) a
lefthanded quark $\qL$ can only annihilate with a righthanded
antiquark $\qbR$ and vice versa.

The diagram of Fig.\ \ref{fig1} is calculated by first evaluating the
amplitude for (\ref{2}) and folding it then with the parton
distributions of the hadrons $h_1,h_2$. In early theoretical work, one
usually assumed that for unpolarised hadrons $h_1,h_2$ the ``parton
beams'' delivered by them are also unpolarised. We will call this the
no-polarisation assumption. In the simplest approximation one
furthermore assumes the partons $q$ and $\bar{q}$ to be strictly
collinear with the hadrons $h_{1,2}$.  This leads to a well known
angular distribution of the lepton pair in the rest frame of the
virtual photon $\gamma^*$. With the polar and azimuthal angles of the
outgoing $l^+$, $\theta$ and $\phi$, one gets
\begin{equation}\label{3}
\frac{1}{\sigma}\frac{d\sigma}{d\Omega}=\frac{3}{16\pi}
\left(1+\cos^2\!\theta\right).
\end{equation}
Here and in Eq.~(\ref{4}) below we use the Collins-Soper reference
frame \cite{3} where the $\gamma^*$ is at rest and the basis vectors
${\mf e}_{1,3}$ are
defined by ${\mf e}_{1,3}=(\widehat{\mf p}_1\pm\widehat{\mf
p}_2) /|\widehat{\mf p}_1\pm\widehat{\mf
p}_2|$, with $\widehat{\mf p}_i={\mf p}_i/|{\mf
p}_i|$. The Collins-Soper frame is 
obtained from the $h_1,h_2$ c.m.\ system by a rotation free boost.

In general, the angular distribution of the $l^+$ is described by three 
functions
$\lambda,\mu,\nu$ which may depend on the kinematic variables of (\ref{1})
\begin{equation}
\label{4}
\frac{1}{\sigma}\frac{d\sigma}{d\Omega}=\frac{3}{4\pi}\,\frac{1}{\lambda+3}
\left(1+\lambda\,\cos^2\!\theta
+\mu\,\sin 2\theta\,\cos\phi
+\frac{\nu}{2}\sin^2\!\theta\,\cos 2\phi \right)\,.
\end{equation}
The LO pQCD result (\ref{3}) implies for the functions $\lambda=1,\:\mu=\nu=0$. Higher
order corrections in $\alpha_s$ change these values. But within the standard
framework and using the no-polarisation ansatz one still finds \cite{4}
at NLO one relation among the coefficients in (\ref{4}):
\begin{equation}\label{5}
1-\lambda-2\nu=0\,.
\end{equation}
This Lam-Tung relation is almost unchanged at NNLO \cite{5}, and even
holds for the inclusion of parton transverse momentum and soft gluon effects
\cite{5a}. However, the relation (\ref{5}) is drastically
violated in experiments \cite{6,7,8}.

In Refs. \cite{10} and \cite{Boer99} two at first sight quite
different ideas have been proposed giving possible explanations for
the violation of the Lam-Tung relation (\ref{5}). It is the purpose of
the present article to give a short review and a detailed comparison
of these two approaches. We also sketch a calculation of instanton
effects for the Drell-Yan reaction and pose some general questions
concerning parton factorisation and entanglement.

\section{Spin effects and factorisation}

In \cite{9} it was argued on general grounds that the assumption
of unpolarised parton beams from a reaction with unpolarised initial hadrons is
questionable due to possible vacuum effects. In particular, it was speculated
that the fluctuating chromomagnetic vacuum fields which are due to the nonperturbative
vacuum structure in QCD might lead to a correlated spin orientation of $q$ and
$\bar{q}$ in (\ref{2}) before the annihilation. This would be in analogy to the
Sokolov-Ternov effect \cite{st}, well known from $e^+e^-$ storage rings. 

In \cite{10} this idea
was expanded upon and confronted with experiments. A general
two-particle spin-density matrix
for the $q\bar{q}$ pair in (\ref{2}) prior to the annihilation was assumed
\begin{eqnarray}\label{rhoqq}
\rho^{(q,\bar{q})}&=&\frac{1}{4}\left\{\mathbbm{1}\otimes \mathbbm{1}+
F_j\,(\mf{\sigma}\cdot \mf{e}_j^{\,*})\otimes \mathbbm{1}\right.		\nonumber\\
&&\left.+G_j\,\mathbbm{1}\otimes(\mf{\sigma}\cdot\mf{e}_j^{\,*})+H_{ij}\,	
(\mf{\sigma}\cdot\mf{e}_i^{\,*})\otimes
(\mf{\sigma}\cdot\mf{e}^{\,*}_j)\right\}\,.
\end{eqnarray}
The quantities $F_i$, $G_i$ and $H_{ij}$ are real functions of the
invariants of the problem. Here we work in the $q\bar{q}$ c.m.\ system and 
set, 
\begin{eqnarray}\label{6a}
\mf{e}^{\,*}_3&=&\frac{ \mf{k}_1^{*}}{|\mf{k}^{*}_1|}\,,\nonumber\\[1ex]
\mf{e}^{\,*}_1&=&\frac{(\mf{p}^{\,*}_1+\mf{p}^{\,*}_2)\times\mf{e}^{\,*}_3}
{|(\mf{p}^{\,*}_1+\mf{p}^{\,*}_2)\times\mf{e}^{\,*}_3|}\,,\nonumber\\[1ex]
\mf{e}^{\,*}_2&=&\mf{e}^{\,*}_3\times\mf{e}^{\,*}_1.
\end{eqnarray}
Such a spin matrix will certainly affect the $\gamma^\ast$-production
cross section from a $q\bar{q}$ state. The related production matrix
in the $q\bar{q}$ c.m.\ system (the angular distribution of $l^+$ arises from the
contraction with the lepton-production matrix) is,
\begin{equation}
\begin{split}
 r_{ij}^{\gamma^\ast}(\mf{k_1^\ast},\rho^{(q,\bar{q})};q\bar{q})&
={\sum_\text{colours,\,spins}}\langle\gamma^\ast_i|\mathcal{T}|q(\mf{k_1^\ast},\alpha,A)\,\bar{q}(-\mf{k_1^\ast},\beta,B)\rangle
\\&\quad\times\tfrac{1}{9}\,\delta_{AA^\prime}\delta_{BB^\prime}\,
\rho^{(q,\bar{q})}_{\alpha\beta\,\alpha^\prime\beta^\prime}
\langle\gamma^\ast_j|\mathcal{T}|q(\mf{k_1^\ast},\alpha^\prime,A^\prime)\,\bar{q}(-\mf{k_1^\ast},\beta^\prime,B^\prime)\rangle^\ast\,.
\label{rij}
\end{split}
\end{equation}
Here $\alpha,\beta,\alpha^\prime,\beta^\prime$ are the spin indices, 
$A,B,A^\prime,B^\prime$ the colour indices and we have assumed no
polarisation in colour space. In LO the amplitude for $\gamma^\ast$
production reads,
\begin{equation}
\begin{split}
 \langle\gamma^\ast_\mu|\mathcal{T}|q(k_1,\alpha,A)\,\bar{q}(k_2,\beta,B)\rangle
&	\equiv\langle 0|e\,J_\mu(0)|q(k_1,\alpha,A)\,\bar{q}(k_2,\beta,B)\rangle \\
&=e\,Q_q\,\delta_{AB}\,\bar{v}_\beta(k_2)\gamma_\mu u_\alpha(k_1)\,.
\label{qbarqtogamma}
\end{split}
\end{equation}
Here $e\,J_\mu$ is the hadronic part of the electromagnetic
current. The conventions for the Dirac spinors are as in \cite{Nachtmann:1990ta} with
$\alpha=\pm 1/2$ ($\beta=\pm 1/2$) representing the quark (antiquark)
with spin orientation in the direction $\pm\mf{e}^{\,*}_3$.

With the standard no-polarisation assumption in spin space, one sets
\begin{equation}\label{6b}
\rho^{(q,\bar{q})}|_\textup{naive}=\tfrac{1}{4}(\mathbbm{1}\otimes\mathbbm{1})
\equiv\tfrac{1}{4}(\delta_{\alpha\alpha^\prime}\,\delta_{\beta\beta^\prime})
\,.
\end{equation}
It was shown in \cite{10} that a nonzero correlation coefficient
\begin{equation}\label{6c}
\kappa \equiv \frac{H_{22}-H_{11}}{1+H_{33}}
\end{equation}
could easily explain the experimentally observed deviation from the Lam-Tung
relation. Indeed, defining
\begin{equation}
 \bar{\kappa}\equiv -\tfrac{1}{4}(1-\lambda-2\nu)\,,
\label{kappabar}
\end{equation}
  one finds instead of (\ref{5}) with (\ref{rhoqq}) and (\ref{6c})
\begin{equation}\label{7}
 \bar{\kappa}\approx\langle\kappa\rangle\,.
\end{equation}
Here the average is over the parton longitudinal and transverse
momenta, see \cite{10}. In Eq.~(\ref{7}) we do not have an equality
sign since higher order perturbative contributions give already a
(very) small contribution to $\bar{\kappa}$ even for $\kappa=0$. 

A good fit to the data of \cite{6,7} could be obtained with the simple ansatz
\begin{equation}\label{8}
\kappa=\kappa_0\,\frac{|\mf{k}_T|^4}{|\mf{k}_T|^4+m^4_T}\,,\hspace{1cm}
\kappa_0=0.17\,,\hspace{0.5cm}m_T=1.5~{\rm GeV}\,,
\end{equation}
where ${\mf k}_T$ is the $\gamma^\ast$ transverse momentum in the
hadronic c.m. system.

In \cite{10} simplifying assumptions for the density matrix
(\ref{rhoqq}) were made	
\begin{equation}
 F_2=F_3=G_2=G_3=H_{12}=H_{13}=H_{21}= H_{31}=0\,.
\label{assumptions}
\end{equation}
This is irrelevant for the Drell-Yan process (\ref{2}). One
can easily show that the $l^+$ angular distribution is only sensitive
to the two parameters, $H_{33}$ and $\kappa$.

Further discussions of possible QCD vacuum effects for the Drell-Yan
and other reactions were given in \cite{bhn,lecture}.

The problem of the angular distribution in the Drell-Yan process was
attacked from a different side in \cite{Boer99}. It was pointed out
that there can be nontrivial spin and transverse momentum correlations
even inside an unpolarised hadron. In the notation of Ref.\
\cite{Boer99} the distribution of quarks (with lightcone momentum fraction
$x_1$ and transverse momentum $\bm{k}_{1T}$) inside an unpolarised 
hadron $h_1$ (with momentum $p_1$ and mass $M_1$) 
is given by a correlation function
$\Phi(x_1,\bm{k}_{1T})$, parametrised as follows:
\begin{equation}
\label{Phi}
\Phi(x_1,\bm{k}_{1T}) =
{f_1(x_1,\bm k_{1T}^2)}\, \frac{\gamma^-}{2} + 
{h_1^\perp (x_1,\bm{k}_{1T}^2)} \, 
\frac{i \slsh{k_{1T}}\,\gamma^-}{2M_1}\,,
\end{equation}
where the transverse momentum is w.r.t.\ the plane spanned by the hadron 
momenta
$p_1$ and $p_2$ in the hadronic centre of mass frame. 
In that frame 
$p_1$ and $p_2$ are predominantly in the lightlike $n_+$ and $n_-$ directions,
respectively. For details we refer to Refs.\ \cite{Boer99,Boer02a}. 

The function $h_1^\perp$ is the distribution of transversely
polarised quarks with nonzero transverse momentum inside an unpolarised
hadron. The subscript $1$ on $f_1$ and $h_1^\perp$ 
indicates that these functions contribute at leading twist and 
should not be confused with the hadron label. The antiquark correlation 
function $\overline{\Phi}(x_2,\bm{k}_{2T})$ is parametrised accordingly:
\begin{equation}
\overline{\Phi}(x_2,\bm{k}_{2T}) =
{\bar{f}{}_1(x_2,\bm k_{2T}^2)}\, \frac{\gamma^+}{2} + 
{\bar{h}{}_1^\perp (x_2,\bm{k}_{2T}^2)} \, 
\frac{i \slsh{k_{2T}}\,\gamma^+}{2M_2}\,.
\end{equation}

The quark spin-density matrix $\rho^{(q)}$ can be
obtained by projecting $\Phi(x_1,\bm{k}_{1T})$ onto the basis
$\left(\psi_{+R},\psi_{+L}\right)$, i.e.\ the right and left chirality
components of the good field $\psi_+ = \frac{1}{2} \gamma^- \gamma^+
\psi$ (see for instance Refs.\ \cite{Bacchetta99,Bacchetta02}); and
analogously for $\overline{\Phi}(x_2,\bm{k}_{2T})$ and the antiquark
spin-density matrix $\rho^{(\bar{q})}$. For given $\bm{k}_{1T}$ and
$\bm{k}_{2T}$ one can boost to the frame (\ref{6a}), which leads
(after appropriate normalisation) to 
\begin{eqnarray}\label{10}
\rho^{(q)}&=&\frac{1}{2}\left\{\mathbbm{1}+\frac{h_1^\perp}{f_1}
\frac{x_1}{M_1} \left(\mf{e}^{\,*}_3 \times \mf{p}_{1}^{\,*} \right) 
\cdot \mf{\sigma} \right\} 
\ \ 
\equiv \ \ 
\frac{1}{2}\left\{\mathbbm{1}+
F_j\,(\mf{\sigma}\cdot \mf{e}_j^{\,*})\right\}\,,
\nonumber\\
\rho^{(\bar{q})}&=&\frac{1}{2}\left\{\mathbbm{1}-
\frac{\bar{h}{}_1^\perp}{\bar{f}{}_1}  
\frac{x_2}{M_2} \left(\mf{e}^{\,*}_3 \times \mf{p}_{2}^{\,*} \right) 
\cdot \mf{\sigma}\right\}\ \ 
\equiv \ \ 
\frac{1}{2}\left\{\mathbbm{1}+
G_j\,(\mf{\sigma}\cdot \mf{e}_j^{\,*}) \right\}\,.
\end{eqnarray}
For simplicity we have suppressed the arguments of the 
functions. 

From Eq.\ (\ref{10}) we arrive at $F_3=0=G_3$ and for $i=1,2$:
\begin{eqnarray}\label{10b}
F_1 = -\frac{h_1^\perp}{f_1} \frac{x_1}{M_1} \mf{p}_{1}^{\,*}
 \cdot \mf{e}^{\,*}_2\,,\hspace{1cm} & & 
F_2 =  +\frac{h_1^\perp}{f_1} \frac{x_1}{M_1} \mf{p}_{1}^{\,*}
 \cdot \mf{e}^{\,*}_1\,, \nn \\
G_1 = + \frac{\bar{h}{}_1^\perp}{\bar{f}{}_1} \frac{x_2}{M_2} 
\mf{p}_{2}^{\,*} \cdot \mf{e}^{\,*}_2\,,\hspace{1cm} & & 
G_2 = - \frac{\bar{h}{}_1^\perp}{\bar{f}{}_1} \frac{x_2}{M_2} 
\mf{p}_{2}^{\,*} \cdot \mf{e}^{\,*}_1\,.
\end{eqnarray}
One observes
that the function $h_1^\perp$ enters in the off-diagonal elements of
$\rho^{(q)}$ and thus corresponds to RL and LR density-matrix elements. 

In the approach followed in Ref.\ \cite{Boer99}, 
the $q\bar{q}$ spin-density matrix is
given by the tensor product of these two nontrivial one-particle spin-density 
matrices,
\begin{equation}\label{11}
\rho^{(q,\bar{q})}=\rho^{(q)}\otimes\rho^{(\bar{q})}\,.
\end{equation}
Clearly, nonzero $h_1^\perp$ implies that the standard no-polarisation ansatz
(\ref{6b}) does not hold. Comparison of Eqs.\ (\ref{11}) and
(\ref{10}) with Eq.\ (\ref{rhoqq}), shows that here $H_{ij} = F_i G_j$
for $i,j=1,2,3$, and hence $H_{i3}=0=H_{3i}$ (due to $F_3=0=G_3$).

The factorisation (\ref{11}) of the spin-density matrix
$\rho^{(q,\bar{q})}$ is usually implicitly assumed once factorisation
of the dependences on hard and soft energy scales is demonstrated for
a process. See for instance Ref.\ \cite{Collins92} for a discussion of
factorisation of the spin-density matrix in the polarised Drell-Yan
process (cf.\ in particular its Eq.\ (14)). For earlier discussions of
issues concerning factorisation for processes where transverse momenta
play a role see e.g.  Refs.~\cite{collins1}. As said, for unpolarised
hadrons it is standard to choose $F_i=0=G_i$. Using instead $F_i$ and
$G_i$ of Eq.\ (\ref{10b}) in a tree level calculation of the Drell-Yan
process leads to $\lambda=1, \mu=0$ and $\nu \neq 0$. The general
expression for $\nu$ in terms of $h_1^\perp$ is given in Ref.\
\cite{Boer99}, but here we will restrict to the case of Gaussian
transverse-momentum dependence for illustration purposes. We assume
that all transverse-momentum-dependent functions are of the form
\begin{equation}
f(x_i,\mf{k}_{iT}^2) = f(x_i) \exp\left(-R^2 \mf{k}_{iT}^2\right)
\frac{R^2}{\pi}\,.
\end{equation}
Moreover, we assume that the width of the Gaussian is the same for $f_1$ and
$\bar{f}{}_1$ (the width will be called $R_f^2$) and similarly for 
$h_1^\perp$ and $\bar{h}{}_1^\perp$ (the width 
will be called $R_h^2$ and should be
larger than $R_f^2$ in order to satisfy a positivity bound). This then leads 
to  	
\begin{equation}
\bar{\kappa} = \frac{\nu}{2} = 
\frac{R_h^2}{4 R_f^2} \; \frac{\bm{k}_T^2}{M_1 M_2} \; 
\exp\left(- \left[ R_h^2-R_f^2 \right] \frac{\bm{k}_T^2}{2} \right)\; 
\frac{\sum_{a} e_a^2 \;h_1^{\perp a}(x_1) h_1^{\perp
\bar{a}}(x_2)}{\sum_{a} e_a^2 \; f_1^a(x_1) f_1^{\bar{a}}(x_2)}\,, 
\label{12}
\end{equation}
where $\bar{\kappa} \equiv -(1-\lambda-2\nu)/4$ and $e_a=e Q_q$, see
Eqs.~(\ref{kappabar}) and (\ref{qbarqtogamma}). We find again that the deviation from the Lam-Tung 
relation arises from an average $\kappa$ (albeit in addition to higher
order perturbative corrections).  In Eq.\ (\ref{12}) $e_a$ denotes the
charge of the quark with flavour $a$; the sum is over flavours and
antiflavours (indicated by $\bar{a}$); and, we have used that
$\bar{f}^a = f^{\bar{a}}$, i.e.\ the distribution of antiquarks of
flavour $\bar{a}$ inside a hadron $h$ is equal to the distribution of
quarks of flavour $a$ inside an anti-hadron $\bar{h}$.
 
Setting $R_f^2 = 1$ GeV$^{-2}$ and fitting the NA10 data \cite{7} as done in 
Ref.\ \cite{Boer99}, leads to a good fit for $R_h^2-R_f^2 = 0.17 R_f^2$ 
and 
\begin{equation} 
\Bigg\langle 
\frac{\sum_{a} e_a^2 \;h_1^{\perp a}(x_1) h_1^{\perp
\bar{a}}(x_2)}{\sum_{a} e_a^2 \; f_1^a(x_1) f_1^{\bar{a}}(x_2)} \Bigg\rangle
= 0.02\,, 
\end{equation}
where we consider the average over $x_1$ and $x_2$.
Assuming $u$-quark dominance and $h_1^\perp/f_1 \approx
\bar{h}{}_1^{\perp}/\bar{f}{}_1$, this leads to the reasonable result that 
on average $h_1^\perp$ is approximately 14\% of the size of $f_1$. 

This result is of course dependent on the assumptions, 
but it serves the purpose of 
illustrating that the data can in principle be explained by a nonzero
$h_1^\perp$. Hence, in order to experimentally 
discriminate between the two approaches of Refs.\ \cite{10,Boer99}, 
more data is clearly needed, either from other kinematic regions or from 
other processes. In the next section we will elaborate on what is required 
and what are the opportunities for distinguishing between the two approaches. 

\section{Comparison of the two approaches}

In this section we compare the approaches of \cite{10} and
\cite{Boer99}. Let us first of all emphasise that the ansatz of
\cite{Boer99}, given by Eqs.~(\ref{10}) to (\ref{11}) is perfectly
compatible with the general ansatz (\ref{rhoqq}) put forward in
\cite{10}, but restricts $\rho^{(q,\bar{q})}$ to be factorising.

There is a further restriction in the ansatz~(\ref{10}) to
(\ref{11}). It requires $F_3=G_3=0$. This comes about as follows.
The correlation function $\Phi$ in (\ref{Phi}) for the hadron $h_1$ is
supposed to depend only on the momenta $p_1, p_2, k_1$ (actually only on the
direction of $p_2$), the
correlation function $\overline{\Phi}$ for $h_2$ only on $p_2, p_1,
k_2$. From three four-vectors we can only form one axial vector in
each case,
\begin{equation}
 a_{1,2}^{\mu}=\epsilon^{\,\mu\nu\rho\sigma}p_{1\nu}\,p_{2\rho}\,k_{1,2\,\sigma}\,.
\label{pseudovec}	
\end{equation}
To form a pseudoscalar invariant we need all four independent
four-vectors
\begin{equation}
 \mathcal{A}=\epsilon^{\,\mu\nu\rho\sigma}p_{1\mu}\,p_{2\nu}\,k_{1\rho}\,k_{2\sigma}\,.
\label{pseudoscalar}
\end{equation}
Now $F_3$ and $G_3$ are in essence measuring the degree of
longitudinal polarisation of the quark $q$ and antiquark
$\bar{q}$. Therefore, due to parity invariance of the strong
interaction, $F_3$ and $G_3$ must be pseudoscalar quantities and thus
linear in $\mathcal{A}$ (\ref{pseudoscalar}). But in the ansatz
(\ref{10b}), (\ref{11}) $F_3$ arises from the correlation function
$\Phi$ of hadron $h_1$, see (\ref{Phi}), and can thus only depend on three four-vectors, from
which we cannot form a pseudoscalar invariant. Thus, with the ansatz
(\ref{10}) to (\ref{11}), $F_3$ must be zero. The same holds for
$G_3$. Therefore, the ansatz of \cite{Boer99} implies $F_3=G_3=0$ and
due to the factorisation of the $q\bar{q}$ matrix $H_{33}=0$. 
In the general ansatz of \cite{10} the density matrix can from the
outset depend on all four four-vectors of the problem, there is a
pseudoscalar invariant (\ref{pseudoscalar}) available, and $F_3$,
$G_3$ do not have to vanish. Obviously, also $H_{33}$ does not need to
be zero in the general approach.

As mentioned in section 2, the Drell-Yan reaction (\ref{2}) is only
sensitive to the density-matrix element $H_{33}$ and the combination
$\kappa$ (see Eq.~(\ref{6c})). Therefore, one way to check if the restricted
form (\ref{11}) of $\rho^{(q\bar{q})}$ is actually realised would be
to measure $H_{33}$. But, as already mentioned in \cite{10}, the
normalised  angular distribution (\ref{4}) of the lepton pair is
practically only sensitive to $\kappa$. A factor of $1+H_{33}$ enters in
the cross section formula but influences mainly the absolute
normalisation. This latter effect is
difficult to measure due to uncertainties in the quark and antiquark
distributions and in higher order contributions giving rise to the so-called 
K-factors. Thus we are left with one relevant parameter $\kappa$. 

Different physical mechanism were proposed in \cite{10} and
\cite{Boer99} to produce a nontrivial $q\bar{q}$ density matrix with
$\kappa\neq 0$. In \cite{10} it was suggested that effects of the
nontrivial QCD vacuum may be responsible for $\kappa\neq 0$. In
\cite{Boer02b,LuMa04} model calculations using the general framework
of \cite{Boer99} were performed showing that initial-state gluon
exchange can produce $\kappa\neq 0$.

Let us see if on general grounds we can expect different behaviour for
the observable quantity $\bar{\kappa}$ (\ref{kappabar}) from these two
physical pictures. One possibility for comparison would be to study
$\bar{\kappa}$ as a function of $\bm{k}_T$. The ansatz given in Eq.\
(\ref{8}) --taken literally-- implies that $\bar{\kappa} \approx
\kappa_0$ for large $\bm{k}_T$. 
This is a very different behaviour than that expected from an underlying
$h_1^\perp$ function, which is assumed to vanish for large quark
transverse momentum, in accordance with the ansatz of factorisation of
hard and soft energy scales in the process. This forces $\bar{\kappa}$ to
vanish (at least at tree level) in the limit of large $|\bm{k}_T|$. Higher
order $\alpha_s$ corrections may modify this conclusion. However, as
mentioned the NNLO corrections were shown to be small \cite{5,10}. Their
(negative) contribution to $\bar{\kappa}$ was found to be well below 1\%
for $|\bm{k}_T|$ values up to 3 GeV (see figure 6 of \cite{10}).
Therefore, one expects $\bar{\kappa}$ (possibly corrected for the small
higher order perturbative contributions) to decrease. A constant
$\bar{\kappa}$, that is both positive and large, for large $\bm{k}_T$
would therefore be irreconcilable with the approach of \cite{Boer99}. In
the general framework of \cite{10} such a behaviour for $\bar{\kappa}$
would be possible but is certainly not required.

The dependence of $\bar{\kappa}$ on the other scale in the process, the
lepton pair invariant mass (denoted by $m_{\gamma^*}$ in Ref.\
\cite{10} and by $Q$ in Ref.\ \cite{Boer99}), may also be different in
the two approaches. Unfortunately it is not clear what would be the generic 
$Q^2$ behaviour of $\bar{\kappa}$ due to vacuum effects.
Regarding $\bar{\kappa}$ arising from nonzero $h_1^\perp$, the
expectation is that it will decrease approximately as $1/Q$ for large
$Q$. This is based on results from Ref.\ \cite{Boer01}, where the
influence of soft gluons on similar azimuthal spin asymmetries was
considered. This means that although $\bar{\kappa}$ is not power-suppressed
at tree level, higher order $\alpha_s$ contributions effectively give
rise to power suppression. Note that this is quite different from
dynamical higher twist contributions, such as discussed in Ref.\
\cite{Brandenburg-Brodsky}, which typically lead to $\nu \sim {\cal O}
(\langle k_T^2 \rangle/Q^2)$ and therefore, are expected to be important only 
for $Q$ values smaller than the
experimentally measured range from 4 GeV up to 12 GeV. 
In any case, the fall-off or persistence of $\bar{\kappa}$ with increasing 
$Q$ could be a discriminating feature, similar to the $\bm{k}_T$ dependence.

A further possibility to differentiate between the two approaches is to 
investigate a possible flavour dependence of $\bar{\kappa}$ 
by varying the types of beams 
($\pi^\pm$, $p$, $\bar{p}$). Clearly vacuum effects do not favour a flavour 
dependence. On the other hand, if the ratio $h_1^\perp/f_1$ varies for 
different flavours and different hadrons, then this could lead to an 
observable flavour dependence. Thus far only $\pi^- \, N$ data have been 
published, although Ref.\ \cite{8} mentions also to have data for a 
$\pi^+$ beam at the same energy. 

Both vacuum effects and nonzero $h_1^\perp$ could lead to a $\bar{\kappa}$
that varies as a function of $(x_1,x_2)$. In addition, observable flavour
dependence of this $x_i$ dependence would arise if the ratio
$h_1^\perp(x_i)/f_1(x_i)$ varies differently as a function of $x$ for
different flavours and different hadrons. This includes the possibility
that $h_1^\perp(x_i)/f_1(x_i)$ changes its sign as a function of $x_i$,
which would lead to sign changes in $\bar{\kappa}$ as a function of $(x_1,x_2)$,
even when restricting to only one particular process.

As a last point in this discussion we mention that since the approach of 
\cite{Boer99} is based on a factorised $q\bar{q}$ spin-density matrix
(\ref{11}), one can test this type of factorisation by measuring 
several related 
processes, such as semi-inclusive deep-inelastic lepton-nucleon scattering, 
where the $h_1^\perp$ function enters in combination with other functions 
\cite{Boer:1997nt,Gamberg}. See also \cite{Ji:2004wu,ma2}. 
In principle one can determine as many observables as unknown
functions in order to extract $h_1^\perp$ and test the consistency of the
factorised approach. Needless to say, this is quite a formidable task, but
outlines of such a scheme have been discussed in some detail 
in Ref.\ \cite{Boer02a}.

Clearly it will also be very interesting to compare the predictions of the
approaches of \cite{10} and \cite{Boer99} for other Drell-Yan type
processes, for instance $Z$ production as well as $\gamma^\ast\:+$ jet
and $Z+\,$ jet production in hadron-hadron collisions. 

\section{Instanton model}

Thus far we have explained in a rather general way, how a violation of
the Lam-Tung relation (\ref{5}) can arise, if the standard ansatz
(\ref{6b}) does not hold. In both approaches \cite{10,Boer99} the
strength of the violation follows directly from a comparison with the
experiment. It would be of great interest to calculate the relevant
parameter describing the asymmetry (namely $\kappa$) in a certain
model. For the approach of Ref.\ \cite{Boer99} this has been done in
Ref.\ \cite{Boer02b,LuMa04} using a spectator model. It was shown
that initial-state gluon exchange could give rise to a nonzero
$h_1^\perp$ and a corresponding $\bar{\kappa}$ for the $p\bar{p}$ and
$\pi^- p$ initiated Drell-Yan process. In this section we want to take
a different approach, namely to outline a model calculation that is
along the lines of Ref.\
\cite{10}.

We already discussed the possibility that in a nontrivial vacuum
the spins of the partons might be correlated. An intriguing possibility
to describe the vacuum structure is given by instantons. Instantons
\cite{bpst} are nonperturbative fluctuations of the gluon fields and 
well known to induce chirality-violating processes, absent in
conventional perturbation theory~\cite{th}. Especially this feature of
instanton-induced processes is a strong motivation to study the role
of instantons as a source of spin correlations. Along similar lines,
various remarkable effects induced by instantons were
investigated. One can find for instance in Ref.~\cite{koch-ssa} an
estimate of certain single spin asymmetries and in Ref.~\cite{koch-ff}
an estimate of the Pauli form factor of the quark. One also expects an
impact of instantons on the question how the proton spin is built up
from the spin and angular momenta of the constituents, see
e.g. \cite{forte} or the review~\cite{koch-rev}. Recently, an estimate
of an azimuthal spin-asymmetry induced by instantons in semi-inclusive
deep-inelastic scattering was presented in Ref.~\cite{shuryak}.

In a rather qualitative way, the Drell-Yan process was already
investigated in an instanton background in \cite{ellis}. There it was
argued that even in the limit of high energy, instantons may lead to
sizable effects, not suppressed by inverse powers of the energy. But it
should be mentioned that spin effects do not play any specific role in
\cite{ellis}.

Here, we want to emphasise that instantons might indeed violate the
naive ansatz (\ref{6b}) via some additional terms to the spin matrix.
The generic instanton-induced chirality-violating process which
contributes in the Drell-Yan case reads for $n_f$ active
flavours (see Eq.~(\ref{2}) for the similar process in usual
perturbation theory),
\begin{equation}
 \qL+\qbL\to\gamma^\ast+(n_f-1)\,\left[\qR+\qbR\right]+n_g\,g\,
 \label{procinst}
\end{equation} 	
and is sketched in Fig.~\ref{fig2}.  The indices indicate the helicity
of the quarks and antiquarks in the process. Of course, the process
with $R$ and $L$ exchanged everywhere --induced by antiinstantons-- 
must also be taken into account. The important point in our approach
is not the significant complication of the final state in
(\ref{procinst}) which contributes to the final state $X$ in
(\ref{1}), but the different helicity structure in the {\em initial}
state.

We mentioned in section 1 that neglecting quark masses in the process
(\ref{2}) only a quark and an antiquark with different helicities
couple to the photon. So one can split the process (\ref{procinst})
into two stages: during the first stage, the quark (or the antiquark)
will change the helicity and afterwards the quark (antiquark) will
interact in the usual way with the antiquark (quark). The final state
will only change the size of the whole instanton contribution but not
the structure of the related spin matrix $\rho^{(q,\bar{q})}$.
\begin{figure}
\centering
\begin{picture}(422,222)(0,0)
\Line(0,215)(0,25)
\Line(410,215)(410,25)
\Text(415,213)[lb]{\text{$2$}}
\BBox(85,110)(180,210)
\BBox(140,70)(180,30)
\Text(160,50)[cc]{\text{\huge$RL$}}
\Text(132,205)[ct]{$\langle\gamma^\ast_\mu|\mathcal{T}|\qR\,\qbL\rangle$}
\Text(20,130)[cb]{$\qL$}
\ArrowLine(10,125)(30,125)
\ArrowLine(30,95)(10,95)
\Text(20,90)[ct]{$\qbL$}
\Line(70,125)(30,125)
\Line(70,95)(30,95)
\Text(110,165)[cb]{$\qR$}
\ArrowLine(70,160)(115,160)
\ArrowLine(115,130)(70,130)
\Text(110,125)[ct]{$\qbL$}
\ArrowArcn(115,145)(15,0,-90)
\ArrowArcn(115,145)(15,90,0)
\Text(150,150)[cb]{$\gamma^\ast$}
\Photon(130,145)(170,145){3}{4}
\ArrowLine(67,96)(103,96)
\ArrowLine(103,90)(67,90)
 \Vertex(80,87){0.3} 
 \Vertex(90,87){0.3} 
 \Vertex(80,85){0.3} 
 \Vertex(90,85){0.3} 
 \Vertex(80,83){0.3} 
 \Vertex(90,83){0.3} 
\ArrowLine(67,80)(103,80)
\ArrowLine(103,74)(67,74)
\Text(108,85)[lc]{$\bigg\}\,n_f-1$}
\Gluon(62,60)(103,60){3}{4}
 \Vertex(80,56){0.3} 
 \Vertex(90,56){0.3} 
 \Vertex(80,54){0.3} 
 \Vertex(90,54){0.3} 
 \Vertex(80,52){0.3} 
 \Vertex(90,52){0.3} 
\Gluon(62,48)(103,48){3}{4}
\Text(108,54)[lc]{$\}\,n_g$}
\GOval(60,110)(80,12)(0){0.9}
\Text(60,110)[cc]{$I$}
\Text(200,110)[cc]{\text{\large$+$}}
\BBox(300,110)(395,210)
\BBox(355,70)(395,30)
\Text(375,50)[cc]{\text{\huge$LR$}}
\Text(347,205)[ct]{$\langle\gamma^\ast_\mu|\mathcal{T}|\qL\,\qbR\rangle$}
\Text(235,130)[cb]{$\qL$}
\ArrowLine(225,125)(245,125)
\ArrowLine(245,95)(225,95)
\Text(235,90)[ct]{$\qbL$}
\Line(285,125)(245,125)
\Line(285,95)(245,95)
\Text(325,165)[cb]{$\qL$}
\ArrowLine(285,160)(330,160)
\ArrowLine(330,130)(285,130)
\Text(325,125)[ct]{$\qbR$}
\ArrowArcn(330,145)(15,0,-90)
\ArrowArcn(330,145)(15,90,0)
\Text(365,150)[cb]{$\gamma^\ast$}
\Photon(345,145)(385,145){3}{4}
\ArrowLine(282,96)(318,96)
\ArrowLine(318,90)(282,90)
 \Vertex(295,87){0.3} 
 \Vertex(305,87){0.3} 
 \Vertex(295,85){0.3} 
 \Vertex(305,85){0.3} 
 \Vertex(295,83){0.3} 
 \Vertex(305,83){0.3} 
\ArrowLine(282,80)(318,80)
\ArrowLine(318,74)(282,74)
\Text(322,85)[lc]{$\bigg\}\,n_f-1$}
\Gluon(277,60)(318,60){3}{4}
 \Vertex(295,56){0.3} 
 \Vertex(305,56){0.3} 
 \Vertex(295,54){0.3} 
 \Vertex(305,54){0.3} 
 \Vertex(295,52){0.3} 
 \Vertex(305,52){0.3} 
\Gluon(277,48)(318,48){3}{4}
\Text(322,54)[lc]{$\}\,n_g$}
\GOval(275,110)(80,12)(0){0.9}
\Text(275,110)[cc]{$I$}
\Text(197,10)[cc]{$=\left|\, a^{(I)}\,\langle\gamma^\ast_\mu|\mathcal{T}|\qR\,\qbL\rangle
\:+\:b^{(I)}\,\langle\gamma^\ast_\mu|\mathcal{T}|\qL\,\qbR\rangle\,\right|^{\,2}$}
\end{picture}
\caption{\small\label{fig2} 
 The instanton-induced process
 $\qL+\qbL\to\gamma^\ast+\mbox{$(n_f-1)\,\left[\qR+\qbR\right]$}+n_g\,g\,.$}
\end{figure}

In Fig.~\ref{fig2} we show the two amplitudes contributing to the
process (\ref{procinst}) and for each amplitude the split into two
stages. In the left part (labelled in Eq.~(\ref{rhoI}) with ($t$)) the
incoming {\em quark} changes the helicity. The right part $(u)$ is of
course similar but the incoming {\em antiquark}	 changes the helicity. It is
sketched that both amplitudes factorise into an instanton part
described by the coefficients $a^{(I)}$ and $b^{(I)}$ and
chirality-conserving amplitudes, namely
$\langle\gamma^\ast_\mu|\mathcal{T}|\qR\,\qbL\rangle$ or
$\langle\gamma^\ast_\mu|\mathcal{T}|\qL\,\qbR\rangle$.

For the simplest instanton-induced process with $n_f=1$ and $n_g=0$
(see \cite{mrs} for a detailed calculation of the related process in
lepton-hadron scattering) one would expect a trivial connection
between $a^{(I)}$ and $b^{(I)}$. For the general case this will change
because of the more complex kinematics, related to the additional
momenta of the final state partons.

The important point is that the two processes shown in Fig.~\ref{fig2}
lead from the same initial to the same final states. Therefore these
amplitudes must be added coherently. This gives in the cross section a
term
\begin{eqnarray}
&&\nn\left(\mathcal{T}^{(t)}_\text{$\mu\,LL$}+
\mathcal{T}^{(u)}_\text{$\mu\,LL$}\right)
\left(\mathcal{T}^{(t)}_\text{$\nu\,LL$}+
\mathcal{T}^{(u)}_\text{$\nu\,LL$}\right)^\ast=
|a^{(I)}|^2\,\langle\gamma^\ast_\mu|\mathcal{T}|\qR\,\qbL\rangle
\langle\gamma^\ast_\nu|\mathcal{T}|\qR\,\qbL\rangle^\ast
\\ &&\nn \qquad
+a^{(I)}\,b^{(I)\,\ast}
\langle\gamma^\ast_\mu|\mathcal{T}|\qR\,\qbL\rangle
\langle\gamma^\ast_\nu|\mathcal{T}|\qL\,\qbR\rangle^\ast
 +
a^{(I)\,\ast}\,b^{(I)}
\langle\gamma^\ast_\mu|\mathcal{T}|\qL\,\qbR\rangle
\langle\
\gamma^\ast_\nu|\mathcal{T}|\qR\,\qbL\rangle^\ast
\\ && \qquad +
|b^{(I)}|^2
\langle\gamma^\ast_\mu|\mathcal{T}|\qL\,\qbR\rangle
\langle\gamma^\ast_\nu|\mathcal{T}|\qL\,\qbR\rangle^\ast\,.
\label{rhoI}
\end{eqnarray}
 Comparing the general amplitude
(\ref{rij}) with the instanton-induced one (\ref{rhoI}), we get
the following expressions for the density-matrix elements (the factor
$1/4$ arises from the averaging over the initial state helicities),
\begin{equation}
 \rho_\text{\tiny$RL\,RL$}^{(I)}=\frac{|a^{(I)}|^2}{4}\,,\quad
 \rho_\text{\tiny$LR\,LR$}^{(I)}=\frac{|b^{(I)}|^2}{4}\,, \quad
 \rho_\text{\tiny$RL\,LR$}^{(I)}=\frac{a^{(I)}\,b^{(I)\,\ast}}{4}\,,\quad
 \rho_\text{\tiny$LR\,RL$}^{(I)}=\frac{a^{(I)\,\ast}\,b^{(I)}}{4}\,.
\end{equation}
For the calculation of the spin matrix, we have to add the
contribution from the usual process without instantons in the
background. The naive expectation related to (\ref{6b}) is
$\left.\rho_\text{\tiny$LR\:LR$}\right|_\text{naive}=\left. \rho_\text{\tiny$RL\:RL$}\right|_\text{naive}=1/4$
and
$\left.\rho_\text{\tiny$RL\:LR$}\right|_\text{naive}=\left.
\rho_\text{\tiny$LR\:RL$}\right|_\text{naive}=0$. Adding this to the instanton-induced
contribution we get finally,
\begin{equation}       
 \kappa=-\frac{\rho_\text{\tiny $RL\,LR$}+\rho_\text{\tiny $LR\,RL$}}{\rho_\text{\tiny $RL\,RL$}+\rho_\text{\tiny $LR\,LR$}}
=-\frac{2\,\text{Re}\lf(a^{(I)}b^{(I)\,\ast}\rg)}{2+|a^{(I)}|^2+|b^{(I)}|^2}\,.
\end{equation}
An estimate of $\kappa$ in the simplest case where $n_f=1$ and $n_g=0$
leads to $\kappa\neq 0$ and we expect the same to be true in
general. 

We want to mention that one can expect contributions to $\kappa$ also
from instanton-induced processes without any additional partons in the
final state. In this case an instanton-antiinstanton pair is located
only on one side of the cut appearing in the contributions to the
cross section (in contrast to the squared amplitudes in
Fig.~\ref{fig2} where one instanton will appear on each side of the
cut). Hence, the quark and the antiquark will change the helicity on
one side of the cut and we will also get off-diagonal contributions.

We summarise: The flipping of the helicity of one quark or antiquark
in the initial state which occurs in the instanton-induced
contribution to the Drell-Yan process should give rise to a nonzero
matrix $H_{ij}$ in (\ref{rhoqq}) and finally to $\kappa$. As already
mentioned, the Drell-Yan process (\ref{2}) is not sensitive to $F_i$
and $G_i$, hence we can not say anything about instanton-induced
contributions to $F_i$ and $G_i$.  A more careful analysis of the
instanton-induced contributions to the Drell-Yan process 
including the complete final state in (\ref{procinst}) is beyond the
scope of the present paper. This and the question  whether the
instanton induced processes lead to a factorising or entangled
$q\bar{q}$ density matrix will be investigated elsewhere.

\section{Summary}
In this paper we have discussed the angular distribution of the lepton
pair in the Drell-Yan process (\ref{1}). We considered the lowest
order reaction (\ref{2}) and studied the influence of the
quark-antiquark spin-density matrix on the lepton's angular
distribution.

It is well known that a trivial spin-density matrix (\ref{6b}) is
disfavoured by experiment (see \cite{6,7,8}).  Experiments are well
described by $q\bar{q}$ spin-density matrices having the coefficient
$\kappa\neq 0$. This can be achieved by a $q\bar{q}$ density matrix
which is factorising into nontrivial $q$ and $\bar{q}$ single-particle
density matrices, as assumed in the ansatz of \cite{Boer99}. The
ansatz of \cite{10} is perfectly compatible with this, but would allow
also for truly entangled $q\bar{q}$ pairs, that is a two-particle
spin-density matrix which cannot be written as a tensor product of
one-particle matrices. We have made a detailed comparison of the
approaches \cite{10} and \cite{Boer99} and have shown how they are
related. We have discussed the underlying physical ideas and have
outlined ways to check these ideas experimentally.

We have discussed instanton effects on the quark-antiquark density
matrix and argued that these could induce spin correlations of the
type indicated by experiments. The question whether instantons lead to
a factorising or an entangled $q\bar{q}$ density matrix will be
studied elsewhere.

We think that it is an important question to follow up how to
determine from experiments the complete $q\bar{q}$ density matrix. For
this other reactions besides the Drell-Yan process clearly are needed.
We have given some discussion of this issue in section 3. It would be
fascinating if the $q\bar{q}$ density matrix turned out to be
entangled. Thus, in this article we want to pose the question: can
there be entanglement at the parton level in hadronic reactions?

\section*{Acknowledgements}
We thank C.~Ewerz, J.~P.~Ma and A.~Ringwald for reading drafts of the
manuscript and for useful discussions.  The research of D.B.~has been
made possible by financial support from the Royal Netherlands Academy
of Arts and Sciences. This research was supported by the German
Bundesministerium f\"ur Bildung und Forschung, project no. 05HAT1VHA/0
and 05HT4VHA/0. A.B. was supported by a Heisenberg grant of the
Deutsche Forschungsgemeinschaft.

\end{document}